\documentclass[12pt]{iopart}

\usepackage{epsfig}
\usepackage{multirow}
\usepackage{iopams}

\begin{document}

\thispagestyle{empty}

\vspace*{-2cm}

\centerline{\large {\bf Effects of bone- and air-tissue inhomogeneities on}} 
\centerline{\large {\bf the dose distributions of the Leksell Gamma Knife$^{\circledR}$}}
\centerline{\large {\bf calculated with PENELOPE}}

\vspace{.3cm}

\begin{center}
{\bf Feras M.O. Al-Dweri$^1$, E. Leticia Rojas$^{1,2}$ and Antonio M. Lallena$^1$}\\
{\small $^1$~Departamento de F\'{\i}sica Moderna, 
Universidad de Granada, E-18071 Granada, Spain.\\
$^2$~Instituto Nacional de Investigaciones Nucleares,
Carretera M\'exico-Toluca, km 36.5,
Ocoyoacac, C.P. 52045, M\'exico.}
\end{center}

\vspace{.2cm}

\small{Monte Carlo simulation with PENELOPE (v.~2003) is applied to calculate
Leksell Gamma Knife$^{\circledR}$ dose distributions for heterogeneous
phantoms. The usual spherical water phantom is modified with a
spherical bone shell simulating the skull and an air-filled cube
simulating the frontal or maxillary sinuses. Different simulations of
the 201 source configuration of the Gamma Knife have been carried out
with a simplified model of the geometry of the source channel of the
Gamma Knife recently tested for both single source and multisource
configurations. The dose distributions determined for heterogeneous
phantoms including the bone- and/or air-tissue interfaces show non
negligible differences with respect to those calculated for a
homogeneous one, mainly when the Gamma Knife isocenter approaches the
separation surfaces. Our findings confirm an important underdosage
($\sim$10\%) nearby the air-tissue interface, in accordance with
previous results obtained with PENELOPE code with a procedure
different to ours. On the other hand, the presence of the spherical
shell simulating the skull produces a few percent underdosage at
the isocenter wherever it is situated.}

\vspace{.3cm}

\section{Introduction}

GammaPlan$^{\circledR}$ (GP) (Elekta 1996) is a computer-based
treatment dose planning system designed to calculate the dose
distributions of the Leksell GammaKnife$^{\circledR}$ (GK) for
stereotactic radiosurgery of certain brain diseases. As almost all
radiosurgery planning systems, GP is quite simplistic. Using a
standard set of beam data, the dose distributions in patients are
calculated by adding those corresponding to each one of the 201 beams
of the GK actually present in each particular treatment (Wu \etal 1990,
Wu 1992). GP assumes homogeneous target media and tissue
heterogeneities are not taken into account (Yu and Sheppard 2003).

However, in stereotactic radiosurgery, Solberg \etal (1998) have
pointed out a remarkable disagreement between Monte Carlo (MC) results
and those predicted by the usual planning systems, in case
inhomogeneous phantoms are considered.

In the investigation of dose perturbations produced by
heterogeneities, MC has showed up as a useful tool, mainly because it
accounts, in an adequate way, for the lack of electron equilibrium
near interfaces. For the GK, Cheung \etal (2001), using the EGS4 MC
code, have found discrepancies up to 25\% in case of extreme
irradiation conditions, mainly near tissue interfaces and dose
edges. This contrasts with the sub-millimeter accuracy with which GK
operates (Elekta 1992).

In this paper we have investigated the effects of bone- and air-tissue
interfaces on dosimetric calculations involving the GK. To simulate
the GK, a simplified geometry model of the source channels is
considered to perform the calculations. This model was proposed in
Al-Dweri \etal (2004) and it is based onto the characteristics shown
by the beams after they pass through the treatment helmets. It has
been shown that the collimation system of each source channel acts as
a ``mathematical collimator'' in which a point source, situated at the
center of the active core of the GK source, emits photons inside the
cone defined by the point source and the output helmet collimators.
If an homogeneous target phantom is considered, this simplified model
of the GK produces doses in agreement with those found if the full
geometry of the source channel is considered, with those calculated by
other authors with various MC codes and with the predictions of GP,
for both a single source (Al-Dweri \etal 2004) and different
multisource (Al-Dweri and Lallena 2004) configurations.

In this work we want to use the simplified geometry model of the GK to
calculate doses in case of heterogeneous target phantoms, including
bone- and air-tissue interfaces.  Simulations have been performed by
using the version 2003 of PENELOPE (Salvat \etal 2003). We compare our
findings for the 201 source configuration with those obtained by
Cheung \etal (2001) with EGS4 and by Moskvin \etal (2004) with
PENELOPE (v.~2001). Different situations of the GK isocenter (both far
and near the interfaces) are considered.

\section{Material and Methods}

\subsection{Leksell Gamma Knife$^{\circledR}$ model}

\begin{figure}[htb]
\begin{center}
\epsfig{figure=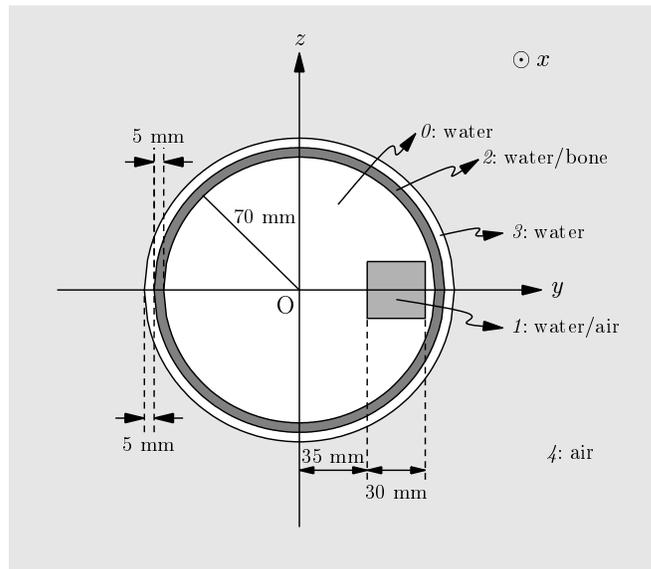,width=9cm}
\end{center}
\vspace*{-0.7cm}
\caption{Schematic plot of the phantom used in our simulations.
\label{fig:phantoms}}
\end{figure}

To study the effect of the heterogeneities, we have used different
configurations of the phantom depicted in figure \ref{fig:phantoms}.
It is chosen to be a sphere with 80~mm of radius made of water except
for the two shadow regions. Region {\it 1}\/ is a cube with a side of
30~mm and with its center at 50~mm of the center of the phantom as
shown in the figure. It is considered to be made of material m$_1$
which can be air (``a''), to simulate the maxillary or frontal
sinuses, or water. Region {\it 2}\/ consists of a 5~mm width spherical
shell with its external surface situated at 5~mm of the phantom
surface. We have considered this shell made of material m$_2$ which
can be either bone (``b''), to simulate the skull, or water
(``w''). The different phantom configurations have been labeled as
${\mathcal P}_{{\rm m}_1 {\rm m}_2}$. With this notation, ${\mathcal
P}_{\rm ww}$ labels the homogeneous phantom. The origin of the coordinate
system is chosen to be at the center of the phantom, as indicated in
the figure. In the figures below, the different regions relevant
to the calculations will be shown in gray scales and labeled with the
corresponding number in italic.

As mentioned above, each one of the 201 sources of the GK are
simulated according the simplified geometry model which is described
in detail in Al-Dweri \etal (2004). It consists of a point source
emitting the initial photons in the cone defined by source itself and
the helmet outer collimators. The coordinates of the 201 point sources
can be found in Al-Dweri and Lallena (2004). They are distributed in
the $z<0$ region and correspond to the situation in which the
isocenter of the GK coincides with the center of the phantom. For the
simulations described below, in which the isocenter is situated at
different positions, these coordinates must be shifted. The position
of the isocenter appears explicitly in the figures as I$[x_{\rm
I},y_{\rm I},z_{\rm I}]$, with the values of the coordinates in mm.

Due to the fact that the distribution of the sources is not completely
uniform, no cylindrical symmetry is shown by the system. Thus, the
doses we have calculated depend on the three cartesian coordinates,
$D^{({\rm m}_1 {\rm m}_2)}(x,y,z)$. The superscript refers to the
materials of the particular phantom ${\mathcal P}_{{\rm m}_1
{\rm m}_2}$ considered in the simulation. Throughout the paper, the
values of the coordinates are given in mm.

Cheung \etal (2001) used a phantom similar to our ${\mathcal P}_{\rm
wb}$. It included the 5~mm width bone shell at 5~mm of the phantom
surface as in our case, but with a full diameter of 180~mm. This
phantom cannot be positioned inside the treatment helmets in such a
way that the isocenter of the GK approaches the skull interface. That
is why in our simulations we have chosen the phantom described above,
which is slightly smaller. In any case, we have compared our results
with those of Cheung \etal by performing simulations with their 
phantom, which we label $\overline{\mathcal P}_{\rm wb}$. To do that
we have calculated the quantity
\begin{equation}
D_{\rm norm}(x,y,z) \, = \,
\displaystyle
\frac{D^{({\rm wb})}(x,y,z)}
{\left[D^{({\rm wb})}(x,y,z) \right]_{\rm max}} \, ,
\label{eq:cheung}
\end{equation}
which corresponds to the dose obtained for 
$\overline{\mathcal P}_{\rm wb}$ divided by its maximum.

A first evaluation of the effects of the different interfaces has been
obtained by calculating the relative differences
\begin{equation}
\label{eq:Delta}
\Delta^{({\rm m}_1 {\rm m}_2)}_{\rm ww}(x_{\rm I},y_{\rm I},z_{\rm I}) 
\, = \, \displaystyle
\frac{D^{({\rm m}_1 {\rm m}_2)}(x_{\rm I},y_{\rm I},z_{\rm I})-
D^{({\rm w} {\rm w})}(x_{\rm I},y_{\rm I},z_{\rm I})}
{D^{({\rm w} {\rm w})}(x_{\rm I},y_{\rm I},z_{\rm I})}
\end{equation}
between the doses obtained at the isocenter for the
heterogeneous, ${\mathcal P}_{{\rm m}_1 {\rm m}_2}$, and homogeneous,
${\mathcal P}_{\rm ww}$, phantoms.

In addition, we have calculated the quantity
\begin{equation}
d^{({\rm m}_1 {\rm m}_2)}_{\rm ww}(x,y,z) \, = \,
\displaystyle
\frac{D^{({\rm m}_1 {\rm m}_2)}(x,y,z)}
{\left[D^{({\rm w} {\rm w})}(x,y,z) \right]_{\rm max}} \, ,
\label{eq:perfil}
\end{equation}
in order to analyze the differences observed in the dose profiles
calculated for the different phantoms.

\subsection{Monte Carlo calculations}

PENELOPE (v.~2003) (Salvat \etal 2003) has been the MC code used to
perform the calculations. PENELOPE permits the simulation of the
coupled transport of electrons and photons, for an energy range from a
few hundred eV up to 1~GeV, for arbitrary materials. PENELOPE provides
an accurate description of the particle transport near interfaces.

\Table{PENELOPE tracking parameters of the materials assumed in our 
simulations. $E_{\rm abs}$($\gamma$) and $E_{\rm
abs}$(e$^{-}$,e$^{+}$) stand for the absorption energies corresponding
to photons and electrons and positrons, respectively.
\label{tab:parameters}}
\br
materials & ~~ & Air & Bone and Water \\ 
\mr
       $E_{\rm abs}$($\gamma$) [keV] && 1.0  & 1.0  \\
$E_{\rm abs}$(e$^{-}$,e$^{+}$) [keV] && 0.1  & 50.0 \\
                             $C_{1}$ && 0.05 & 0.1  \\
                             $C_{2}$ && 0.05 & 0.05 \\
                  $W_{\rm cc}$ [keV] && 5.0 & 5.0 \\
                  $W_{\rm cr}$ [keV] && 1.0 & 1.0 \\
                    $s_{\max}$ [cm]  && $10^{35}$ & $10^{35}$ \\
\br
\end{tabular}
\end{indented}
\end{table}                               

Photons are simulated in PENELOPE in a detailed way. Electrons and
positrons are simulated by means of a mixed scheme which includes two
types of events: hard events, which are simulated in detail and are
characterized by polar angular deflections or energy losses larger
than certain cutoff values, and soft events, which are described in
terms of a condensed simulation based on a multiple scattering theory
(Salvat \etal 2003). The tracking is controlled by means of the four
parameters $C_1$, $C_2$, $W_{\rm cc}$ and $W_{\rm cr}$, as well as the
absorption energies. All these parameters must be fixed for the
materials present in the geometry considered in the simulation. Table
\ref{tab:parameters} shows the values we have assumed in our
simulations. In addition we have fixed the parameter
$s_{\max}=10^{35}$ in all the simulations performed.

The initial source was selected by sampling uniformly between the 201
sources. Initial photons were emitted with the average energy 1.25~MeV
and uniformly in the corresponding emission cone. 

\Table{Composition of the materials assumed in the MC simulations
performed in this work. The values correspond to the weight fraction
of each element in the material. Also the densities are quoted. The
three materials have been generated with the code ~{\tt material}~
included in the PENELOPE package and correspond to the numbers 104,
119 and 277 respectively in the material database of the MC code.
\label{tab:materials}}
\br
& \centre{1}{Air} & \centre{1}{Bone} & \centre{1}{Water} \\
\mr
H  &          & 0.52790 & 0.111894 \\
C  & 0.000124 & 0.19247 &          \\
N  & 0.755267 & 0.01603 &          \\
O  & 0.231781 & 0.21311 & 0.888106 \\
Mg &          & 0.00068 &          \\
P  &          & 0.01879 &          \\
S  &          & 0.00052 &          \\
Ar & 0.012827 &         &          \\
Ca &          & 0.03050 &          \\
\mr
density [g cm$^{-3}$] & 0.0012048 & 1.85 & 1.0  \\
\br
\end{tabular}
\end{indented}
\end{table}

The simulation geometry has been described by means of the geometrical
package PENGEOM of PENELOPE. The three materials assumed in the
simulations performed (air, bone and water) have been
generated with the code ~{\tt material}~ included in the PENELOPE
package. Table \ref{tab:materials} gives the
composition and densities of these materials.

To score the doses we have considered cubic voxels with $\Delta
x=\Delta y=\Delta z=1$~mm, for the 18 and 14~mm helmets, and $\Delta
x=\Delta y=\Delta z=0$.5~mm, for the 8 and 4~mm ones. In the
calculation of the doses $D^{({\rm m}_1 {\rm m}_2)}(x_{\rm I},y_{\rm
I},z_{\rm I})$, used to determine the relative differences
$\Delta^{({\rm m}_1 {\rm m}_2)}_{\rm ww}(x_{\rm I},y_{\rm I},z_{\rm
I})$ as given by equation (\ref{eq:Delta}), voxels with double width
in the $y$ direction were considered. 

The number of histories followed in each simulation has been
$3\cdot10^8$. This permitted to maintain the statistical uncertainties
under reasonable levels. The uncertainties given throughout the paper
correspond to 1$\sigma$. In much of the figures, the error bars do not
show up because they are smaller than the symbols used.

\section{Results}

\subsection{Comparison with EGS4 calculations}

\begin{figure}[htb]
\begin{center}
\hspace*{2cm}
\epsfig{figure=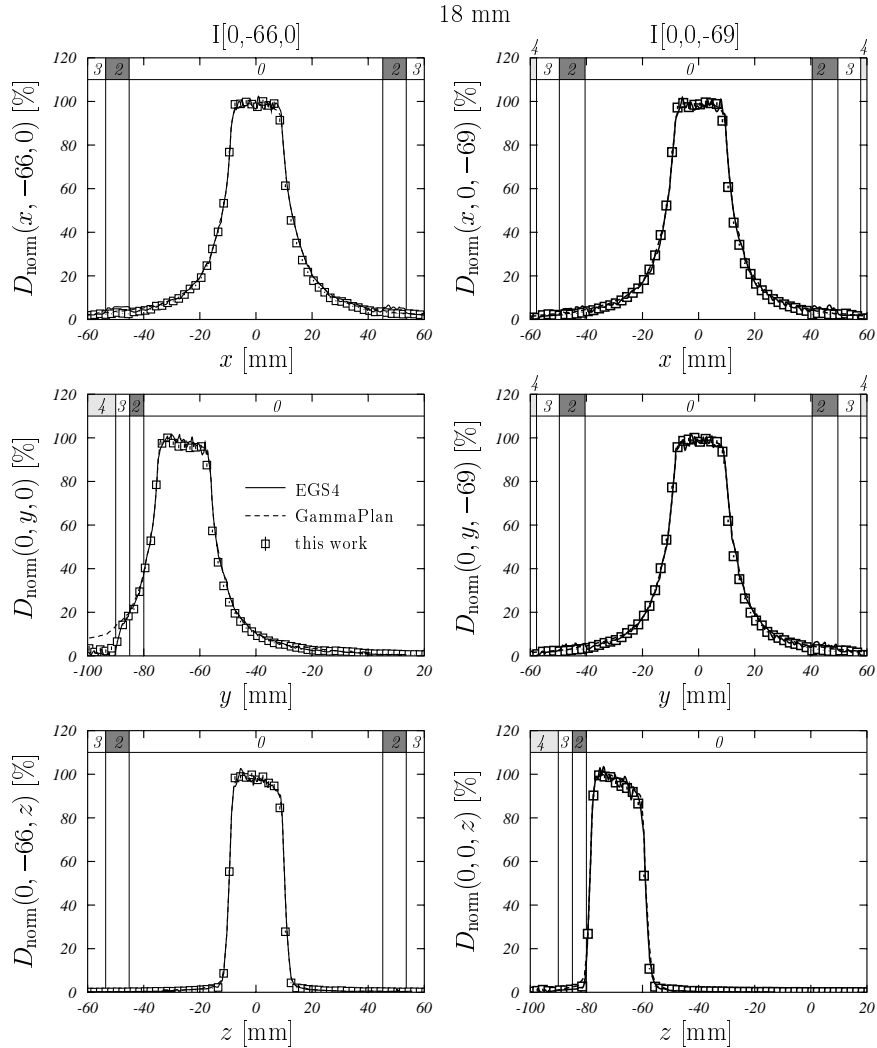,width=12cm}
\end{center}
\vspace*{-0.9cm}
\caption{Dose profiles at the isocenter, relative to their respective
maxima, in percentage, for the 18~mm helmet. Panels show the results
along $x$ (upper), $y$ (medium) and $z$ (lower) axes. Left panels
correspond to the isocenter situated at I[0,-66,0], while
in the right panels the isocenter is at I[0,0,-69]. Open squares are
the results of our simulations. Full curves correspond to EGS4 results
quoted Cheung \etal (2001). Dashed curves correspond to the
predictions of GP quoted by the same authors. The phantom
$\overline{\mathcal P}_{\rm wb}$ has been considered in this case.
\label{fig:cheung}}
\end{figure}

First of all we have compared our results with those obtained by
Cheung \etal (2001) using the EGS4 code. They considered the 18~mm
helmet for two situations of the isocenter: I[0,-66,0] and
I[0,0,-69]. Their results (solid curves) are compared with our
findings (open squares) in figure \ref{fig:cheung}, where the values
of $D_{\rm norm}(x,y,z)$, as given by equation (\ref{eq:cheung}), are
plotted for the three cartesian axes and the two positions of the
isocenter. As we can see, both calculations are in good agreement for
the two cases considered.

In the figure, also the predictions of the GP, quoted by Cheung \etal
(2001), are included (dashed curves). For the isocenter at I[0,-66,0],
a discrepancy between MC simulations and GP results is observed in the
far negative $y$ region (see medium left panel). This is due to the
fact that the GP does not take into account the interfaces and assumes
all tissue to be uniformly represented by water. The same situation is
not observed when the isocenter is at I[0,0,-69] (see right lower
panel), because in that case the dose is roughly zero before reaching
the interface (for $z \sim -80$~mm).

\begin{figure}[htb]
\begin{center}
\epsfig{figure=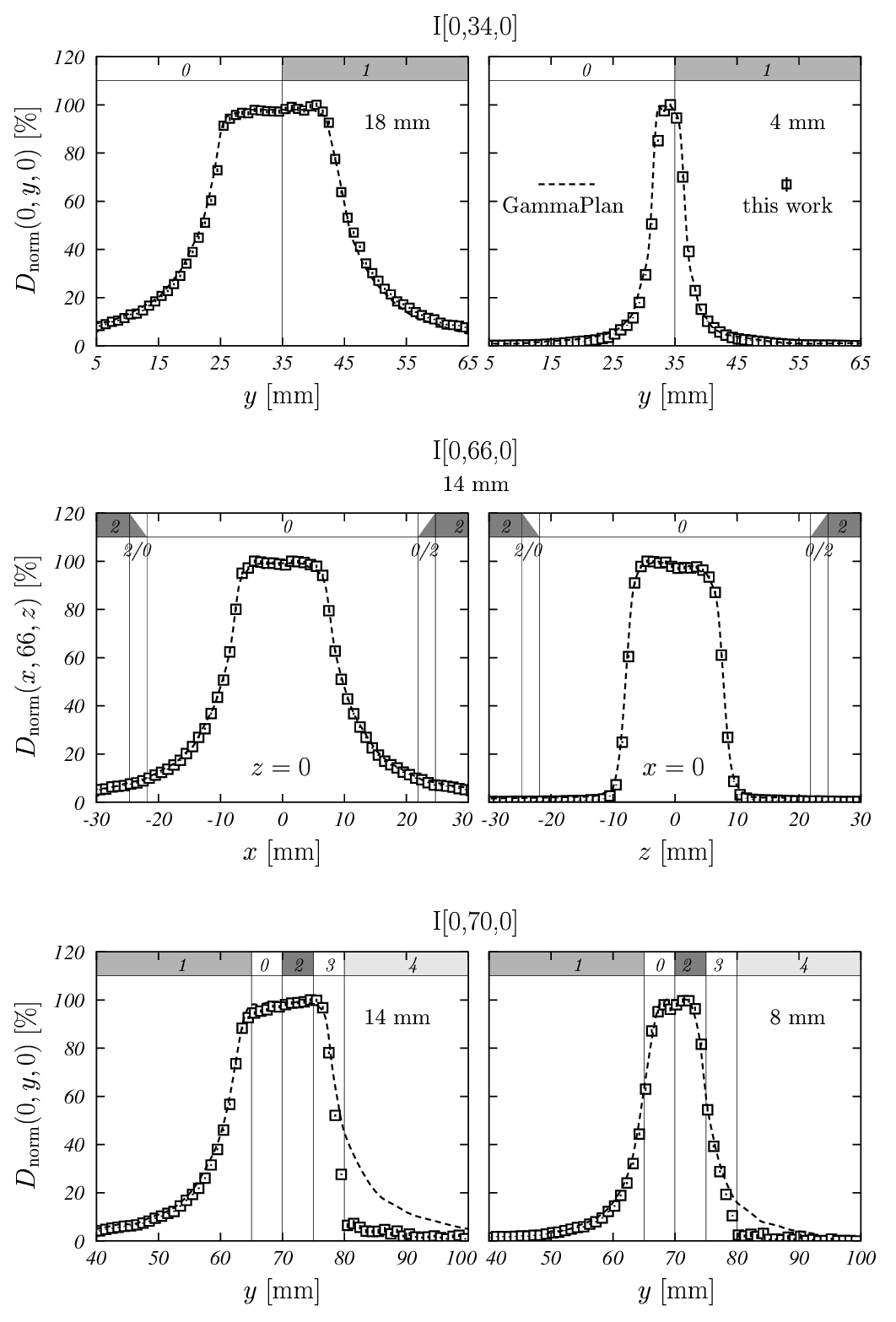,width=10cm}
\end{center}
\vspace*{-0.9cm}
\caption{Dose profiles at the isocenter, relative to their respective
maxima, in percentage, for different positions of the isocenter. Open
squares are the results of our simulations performed with the phantom
$\overline{\mathcal P}_{\rm ww}$. Dashed curves are GammaPlan
predictions (Hamad and Mherat, 2005). Upper panels correspond to the
isocenter situated at I[0,34,0] and show the profiles along $y$ axis
for the 18 and 4 mm helmets. Medium panels show the profiles along $x$
(left) and $z$ (right) axes for the 14 mm helmet with the isocenter
situated at I[0,66,0]. Lower panels are similar to the upper ones but
for the isocenter situated at I[0,70,0] and for the 14 and 8 mm
helmets.
\label{fig:GP}}
\end{figure}

\subsection{Comparison with GammaPlan predictions}
\label{sec:GP}

Figure \ref{fig:GP} shows a comparison of different simulations
performed for the phantom $\overline{\mathcal P}_{\rm ww}$ (open
squares) with GP predictions of Hamad and Mherat (2005) (dashed
curves). In the upper panels the isocenter is situated at I[0,34,0]
and the profiles along $y$ axis are shown for the 18 and 4 mm
helmets. The isocenter situated at I[0,66,0] in medium panels where
the profiles along $x$ and $z$ axes are shown for the 14 mm
helmet. Finally, the profiles along $y$ axis are shown in lower
panels, for the 14 and 8 mm helmets and the isocenter situated at
I[0,70,0].

As we can see, the simulation for the water phantom produces results
in very good agreement with the GP predictions. Below (see section
\ref{sec:DP}), the effects of the interfaces in these cases will be
analyzed and it will be clear that GP cannot describe these effects.

This is evident also in right tail of the profiles shown in the lower
panels. We can see how a clear discrepancy between the simulation and
the GP appears at $y=80$~mm, that is, at the external border of the
phantom. There in, an interface air-water is considered in the
simulation, while GP does not take into account such a situation.

\subsection{Effects of the tissue inhomogeneities on dose at the isocenter}

Now we analyze the results obtained for different positions of the
isocenter of the GK, paying special attention to the situations in
which the isocenter is close to the interfaces.

\begin{figure}[htb]
\begin{center}
\epsfig{figure=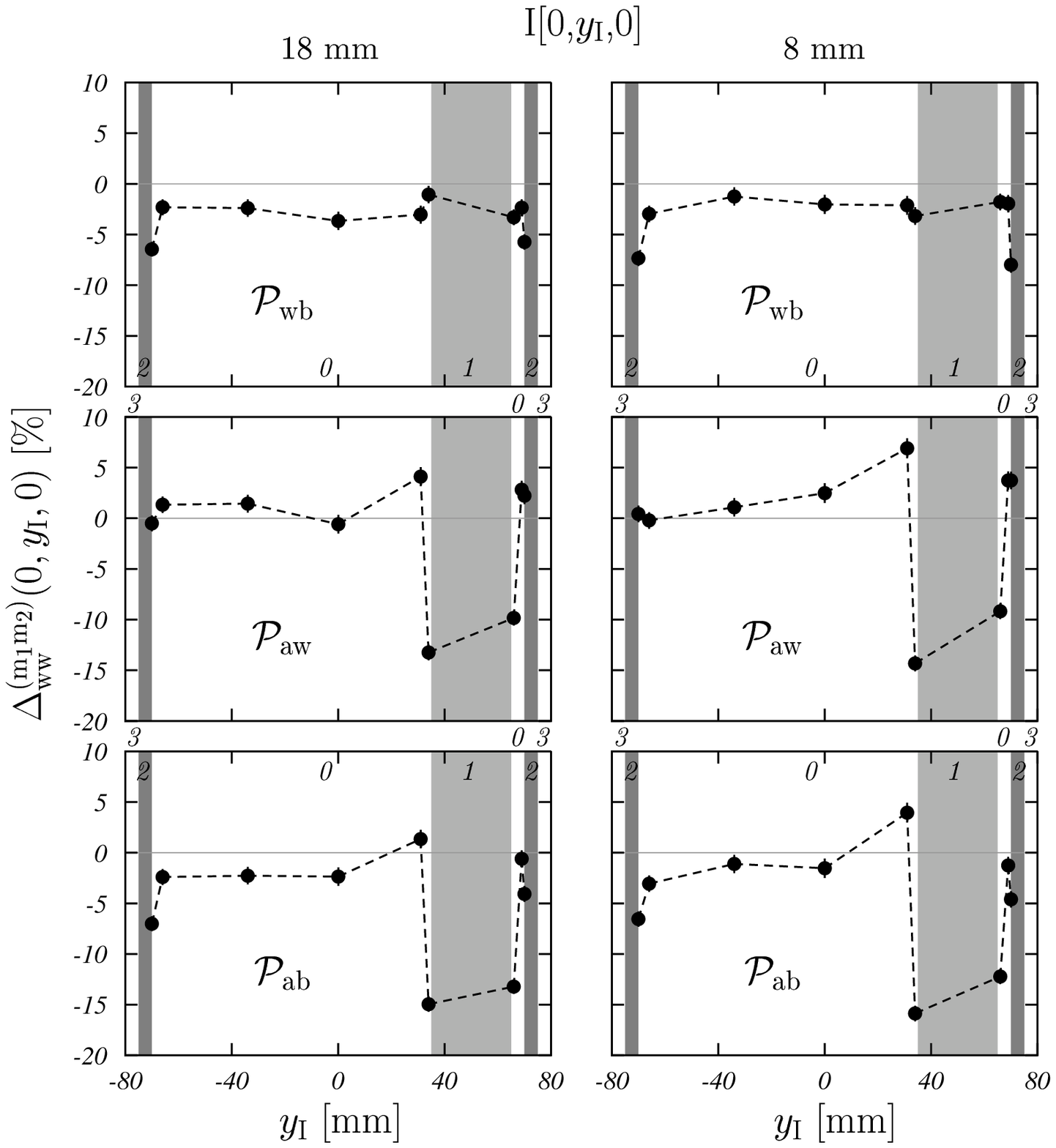,width=10cm}
\end{center}
\vspace*{-0.9cm}
\caption{Relative differences $\Delta^{({\rm m}_1 {\rm m}_2)}_{\rm
ww}(0,y_{\rm I},0)$ in percentage, between the doses calculated at the
isocenter for heterogeneous and homogeneous phantoms (see equation
(\protect\ref{eq:Delta})). Upper panel corresponds to the phantom
${\mathcal P}_{{\rm w} {\rm b}}$; medium panel represents the results
in case the heterogeneous phantom ${\mathcal P}_{{\rm a} {\rm w}}$, and
in the lower panels the phantom ${\mathcal P}_{{\rm a} {\rm b}}$ has
been considered. Results for the 18~mm (left panels) and 8~mm (right
panels) helmets are shown.
\label{fig:deltas}
}
\end{figure}

First, we have investigated the effects of tissue inhomogeneities on
the doses calculated at the isocenter. We have varied its position by
fixing the coordinate $y_{\rm I}$ at different values ranging from
-70~mm to 70~mm and maintaining $x_{\rm I}=z_{\rm I}=0$. The results
obtained for the relative differences with respect to the homogeneous
phantom, $\Delta^{({\rm m}_1 {\rm m}_2)}_{\rm ww}(0,y_{\rm I},0)$, as
given by equation (\ref{eq:Delta}), are shown in figure
\ref{fig:deltas}. Therein, upper panels correspond to the phantom
${\mathcal P}_{{\rm w} {\rm b}}$; medium panels represent the results
in case of the heterogeneous phantom ${\mathcal P}_{{\rm a} {\rm w}}$,
and, finally, in the lower panels the phantom ${\mathcal P}_{{\rm a}
{\rm b}}$ has been considered. We have plotted the results for the
18~mm (left panels) and 8~mm (right panels) helmets. Similar results
are obtained for the 14 and 4~mm helmets.

As we can see, the presence of the bone spherical shell (upper panels)
produces a reduction of the dose at the isocenter with respect to that
obtained for the homogeneous phantom. This reduction is observed at
practically any position of the isocenter, being $\sim$ 3\% for the
two helmets. A higher perturbation in the dose is observed when the
isocenter is situated exactly at the interface bone-water, $y_{\rm I}=
\pm 70$~mm. In this case, the reduction in the dose rises to 5\% for
the 18~mm helmet and it is even larger for the 8~mm one. It seems
evident from our results that the effect of the full skull, which we
simulate here by means of the bone spherical shell, is not negligible
at all.

To have an idea about the origin of this effect, we have evaluated the
reduction in the dose due to the bone inhomogeneity in a very simple
case. We have considered a photon pencil beam coming from the source
and reaching the phantom and we have calculated the dose at the
isocenter neglecting scattering photons. In case the phantom ${\mathcal
P}_{\rm wb}$ is considered, the dose at the isocenter is proportional
to (see e.g. Berger, 1968) 
\begin{equation}
D^{({\rm w} {\rm b})}(x_{\rm I},y_{\rm
I},z_{\rm I}) \propto \displaystyle \left(\frac{\mu_{\rm
en}}{\rho} \right)_{\rm w} \, \frac{E_0}{4\pi r^2} \, \exp \left[
-\mu_{\rm a} s_4 -\mu_{\rm w} s_3 
-\mu_{\rm b} s_2 -\mu_{\rm w} s_0 \right] \, , 
\end{equation}
where $(\mu_{\rm en}/\rho)_{\rm w}$ is the mass energy absorption
coefficient of water at the initial energy of the photons, $E_0$, and
$\mu_{\rm a}$, $\mu_{\rm w}$ and $\mu_{\rm b}$ are the attenuation
coefficients of air, water and bone, respectively, at the same
energy. The values $s_i$ correspond to the length of the trajectory
segments traveled in the region $i$ of the phantom and thus,
\begin{equation}
r \, = \, s_4 \, + \, s_3 \, + \, s_2 \, + \, s_0 
\end{equation}
is the distance from the source to the isocenter. If we consider the
phantom ${\mathcal P}_{\rm ww}$,
\begin{equation}
D^{({\rm w} {\rm w})}(x_{\rm I},y_{\rm
I},z_{\rm I}) \propto \displaystyle \left(\frac{\mu_{\rm
en}}{\rho} \right)_{\rm w} \, \frac{E_0}{4\pi r^2} \, \exp \left[
-\mu_{\rm a} s_4 -\mu_{\rm w} (s_3+s_2+s_0) \right] \, .
\end{equation}
We are interested in the fraction of both doses, which is given by
\begin{equation}
\displaystyle 
\frac
{D^{({\rm w} {\rm b})}(x_{\rm I},y_{\rm I},z_{\rm I})}
{D^{({\rm w} {\rm w})}(x_{\rm I},y_{\rm I},z_{\rm I})} \, = \,
\exp \left[ -(\mu_{\rm b} - \mu_{\rm w}) s_2 \right] \, .
\end{equation}
For the photon energy considered in our simulations, $E_0=1.25$~MeV,
the attenuation coefficients can be calculated easily (see Hubbell and
Seltzer, 2004) and one obtains $\mu_{\rm b}=0.11174$~cm$^{-1}$ and
$\mu_{\rm w}=0.06323$~cm$^{-1}$. On the other hand the length $s_2$
can vary from source to source, depending on the position of the
isocenter. If the phantom is centered with respect to the helmet, that
is if the isocenter is at I[0,0,0], $s_2=5$~mm for all the sources. In
this case the dose ratio is 0.976, and a reduction of 2.4\% is
found. This is the minimum reduction found for all positions of the
isocenter. By varying them in the interval ($-70$~mm,70~mm) in the
three directions, we sample the full volume of the phantom, $s_2$
ranges between 0.5~cm and 2.7~cm and the reduction due to the bone
inhomogeneity varies between 2.4\% and 12.2\%. These results indicate
that, as we have found in our simulations, a few percent reduction in
the dose at the isocenter is expected due to the bone inhomogeneity,
independently of the position of the isocenter.

The air-tissue interface (central panels) produces a slight increase
(1-2\% at most) in the dose at the isocenter, in comparison with that
found for the homogeneous phantom, when it is situated far from the
separation surface. When the interface is approached the relative
difference $\Delta^{({\rm m}_1 {\rm m}_2)}_{\rm ww}$ increases, the
dose at the isocenter for the inhomogeneous phantom is $\sim$5\%
larger than that obtained for the homogeneous phantom and this occurs
until a point very close to separation surface is reached. Once the
isocenter is situated at this position, the dose calculated for the
heterogeneous phantom reduces strongly with respect to that of the
homogeneous one. This reduction is $\sim$ 15\% in the inner side and
$\sim$ 10\% in the outer side of the air cube.

A similar situation is observed when both interfaces are present
(lower panels) except for a general decrease of the dose at the
isocenter whatever $y_{\rm I}$ is. This shift toward smaller doses is
due, as said before, to the presence of the skull, as we can see
clearly in the region $y_{\rm I}<0$, far from the air interface, which
shows a behavior rather similar to that plotted in the upper panels.
One should point out the fact that the overdosage observed in the
region around $y_{\rm I}=30$~mm for the ${\mathcal P}_{{\rm a} {\rm
w}}$ phantom is largely reduced when, in addition, the bone shell is
considered.

\begin{figure}[htb]
\begin{center}
\epsfig{figure=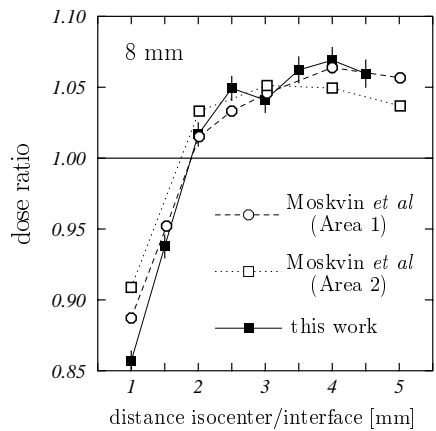,width=6.5cm}
\end{center}
\vspace*{-0.9cm}
\caption{Dose ratio at the isocenter for the doses calculated with the
${\mathcal P}_{\rm aw}$ and the homogeneous ${\mathcal P}_{\rm ww}$
phantoms, as a function of the distance of the isocenter to the
air-tissue interface and for the 8~mm helmet. Black squares are our
results. Open squares and circles are those of Moskvin \etal (2004)
for two different areas of their phantom.
\label{fig:moskvin}
}
\end{figure}

The results obtained near the air-tissue interfaces can be compared
with those found by Moskvin \etal (2004). In figure \ref{fig:moskvin},
we show the dose ratio at the isocenter for the doses calculated with
the ${\mathcal P}_{\rm aw}$ and the homogeneous ${\mathcal P}_{\rm
ww}$ phantoms, as a function of the distance of the isocenter to the
air-tissue interface. Results correspond to the 8~mm helmet. Black
squares are our results. Open squares and circles are those of Moskvin
\etal (2004) for two different areas of their phantom. As we can see,
agreement between the different calculations is rather reasonable.

\subsection{Effects of the tissue inhomogeneities on dose profiles}
\label{sec:DP}

\begin{figure}[htb]
\begin{center}
\hspace*{1cm}
\epsfig{figure=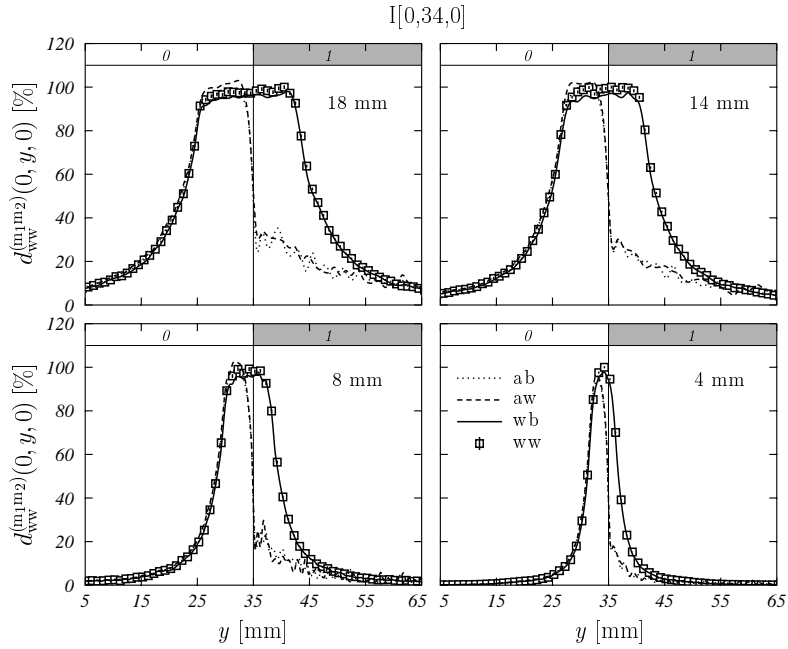,width=11cm}
\end{center}
\vspace*{-.9cm}
\caption{Values of $d^{({\rm m}_1 {\rm m}_2)}_{\rm ww}(0,y,0)$ in
percentage, as given by equation (\protect\ref{eq:perfil}), when the
isocenter is situated at I[0,34,0]. The profiles in the
$y$-axis are shown for the four helmets. The squares correspond to the
homogeneous phantom ${\mathcal P}_{{\rm w} {\rm w}}$. Solid curves
have been obtained with the phantom ${\mathcal P}_{{\rm w} {\rm b}}$.
Dashed curves represent the results in case of the heterogeneous
phantom ${\mathcal P}_{{\rm a} {\rm w}}$. Dotted curves take into
account the phantom ${\mathcal P}_{{\rm a} {\rm b}}$.
\label{fig:perfil-34}
}
\end{figure}

\begin{figure}[htb]
\begin{center}
\epsfig{figure=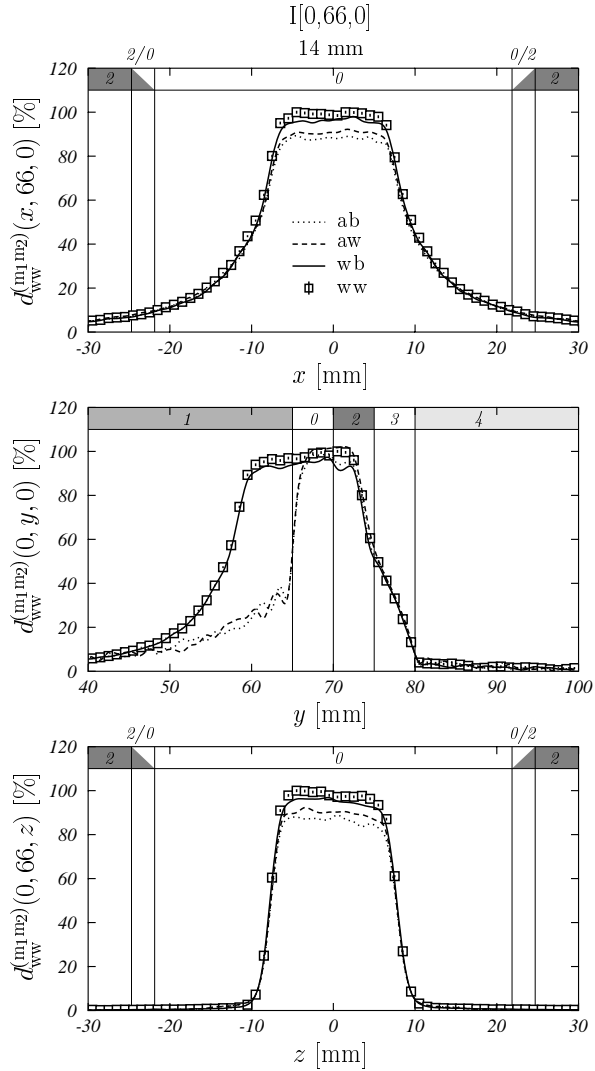,width=8cm}
\end{center}
\vspace*{-.9cm}
\caption{Values of $d^{({\rm m}_1 {\rm m}_2)}_{\rm ww}(x,y,z)$ in
percentage, as given by equation (\protect\ref{eq:perfil}), when the
isocenter is situated at I[0,66,0]. The profiles in the
three cartesian axes are shown for the 14~mm helmet. The lines have
the same meaning as in figure \protect\ref{fig:perfil-34}.
\label{fig:perfil-66}
}
\end{figure}

The larger effects observed appear when the isocenter is situated
nearby an air-tissue heterogeneity. In order to analyze in detail the
dose in this situation, we have calculated the quantities $d^{({\rm
m}_1 {\rm m}_2)}_{\rm ww}(x,y,z)$, as given by equation
(\ref{eq:perfil}), for two positions of the isocenter: I[0,34,0] and
I[0,66,0]. In these two positions the isocenter is at 1~mm distance
from the inner and outer sides of the air cube (region {\it 1}),
respectively. Some results are plotted in figures \ref{fig:perfil-34}
and \ref{fig:perfil-66}.

Figure \ref{fig:perfil-34} shows the profiles along the $y$-axis for
the four helmets and for the isocenter at I[0,34,0]. Therein the
squares correspond to the homogeneous phantom ${\mathcal P}_{{\rm w}
{\rm w}}$, while solid, dashed and dotted curves have been obtained
with the phantoms ${\mathcal P}_{{\rm w} {\rm b}}$, ${\mathcal
P}_{{\rm a} {\rm w}}$ and ${\mathcal P}_{{\rm a} {\rm b}}$,
respectively. If only the bone is considered (solid curves), a
reduction in the plateau region including the maximum dose is
observed. This is the same reduction previously discussed for the dose
at the isocenter.

On the contrary, the presence of the air-tissue interface (dashed
curves) produces a strong reduction of the dose on the ``air'' side
(the right side in this case) of the interface and an enhancement of
the dose profile on the ``water'' side (the left side in this case) of
the separation surface. This effects are better seen for the 18
helmet. The main effect of the simultaneous consideration of both
interfaces (dotted curves) is to cancel the overdosage on the left
side of the interface. The results here obtained are very similar to
those plotted in the figure 6 of the work of Moskvin \etal
(2004). These large differences in the dose produced by the air-tissue
heterogeneities cannot be neglected.

Figure \ref{fig:perfil-66} depicts the results obtained for the
14~mm helmet when the isocenter is situated at I[0,66,0]. The
different curves correspond to the same phantoms as in the previous
figure. Here the profiles along the three cartesian axes are plotted.
Two facts deserve a comment. First, it is appreciable in the upper and
lower panels the reduction of the dose produced by the presence of the
air-tissue interface (dashed and dotted curves). Also, the comparison
between both curves gives us an idea of the additional diminution
produced by the bone shell. Second, it is again evident (see medium
panel) the strong overdosage produced if the air-tissue interface is
not taken into account, but it is also remarkable the reduction in the
dose observed in the region {\it 2}\/ when the bone shell is considered
(solid curve).

\begin{figure}[htb]
\begin{center}
\hspace*{1cm}
\epsfig{figure=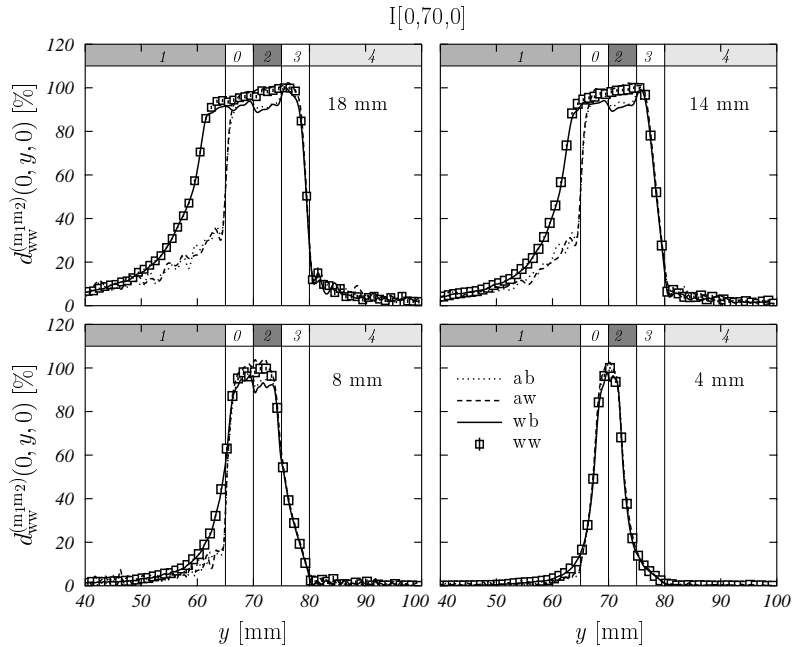,width=11cm}
\end{center}
\vspace*{-.9cm}
\caption{Same as in figure \protect\ref{fig:perfil-34} but for the
isocenter situated at I[0,70,0].
\label{fig:perfil-70}
}
\end{figure}

To complete our analysis, we show in figure \ref{fig:perfil-70}
results similar to those in figure \ref{fig:perfil-34} but for the
isocenter situated at I[0,70,0], that is exactly at the bone-tissue
interface. Apart from the reductions observed in the dose nearby the
air-tissue interface, here the effect of the bone shell in region {\it
2}\/ is, as expected, stronger than in the previous case.

To finish we point out that, as it is observed in figures
\ref{fig:perfil-34} and \ref{fig:perfil-70} and in the medium panel of
figure \ref{fig:perfil-66}, the dose in the interface air-water is, in
all cases, $\sim$50\% of the maximum dose obtained for the homogeneous
phantom. This result is into agreement with the findings of Moskvin
\etal (2004).

\section{Conclusions}

In this work we have investigated the dosimetry of the GK in case of
heterogeneous phantoms by considering a simplified source model for the
single source channels. Calculations have been done by using the Monte
Carlo code PENELOPE (v.~2003) for the configuration including 201
unplugged sources and for different positions of the isocenter of the
GK.

The use of the simplified model produces results for the dose profiles
at the isocenter which are in agreement with previous calculations
done with EGS4, whereas they show discrepancies with the predictions
of the GP, mainly at the interfaces.

In general we can say that the presence of typical tissue
inhomogeneities produces an underdosage with respect to the results
obtained when an homogeneous phantom is considered. This happens for
almost all the positions of the isocenter of the GK. This underdosage
can reach values larger than 10\% in the vicinity of air-tissue
interfaces. The only exception of this conclusion occurs when the
isocenter is situated at a distance of a few millimeters of the
air-tissue separation surface, where an overdosage is produced.
However, this overdosage is very small if, in addition to the tissue
inhomogeneity, also the bone inhomogeneity is considered.

We have analyzed the doses deposited in phantoms including bone and
air inhomogeneities and we have found non-negligible discrepancies
with the doses obtained in case of water homogeneous phantom. In this
respect it is worth to mention that an air inhomogeneity simulating
the maxillary or frontal sinuses, give rise to large modifications of
the dose profiles.

We have found a reasonable agreement with previous calculations
performed by Moskvin \etal with PENELOPE code in the case of the
air-tissue interfaces.

In what refers to the bone-tissue inhomogeneity representing the skull, our
results show a $\sim 3$\% underdosage at the isocenter, with respect
to the doses calculated for the homogeneous phantom. This effect can be
observed wherever the isocenter is situated.

The discrepancies observed between the results obtained for
heterogeneous and homogeneous phantoms suggest that GP predictions
must be corrected in order to take care of the air- and bone-tissues
inhomogeneities, mainly in those cases in which the interfaces are
present nearby the target area.\\

\noindent
{\bf Acknowledgments}\\
{\small We kindly acknowledge A. Hamad and H. Mherat for providing us
  with the GammaPlan predictions quoted in section \ref{sec:GP}.
  F.M.O. A.-D. acknowledges the A.E.C.I. (Spain), the University of
  Granada and the I.A.E.A. for funding his research stay in Granada
  (Spain). E.L.R. acknowledges the University of Granada and the
  Departamento de F\'{\i}sica Moderna for partially funding her stay
  in Granada (Spain). This work has been supported in part by the
  Junta de Andaluc\'{\i}a (FQM0220).}

\References

\item[] Al-Dweri F M O, Lallena A M 2004b 
A simplified model of the source channel of the Leksell Gamma
Knife$^{\circledR}$: testing multisource configurations with PENELOPE
{\it Phys. Med. Biol.}  {\bf 49} 3441-53

\item[] Al-Dweri F M O, Lallena A M and Vilches M 2004a
A simplified model of the source channel of the Leksell
GammaKnife$^{\circledR}$ tested with PENELOPE 
{\it Phys. Med. Biol.} {\bf 49} 2687-703

\item[] Berger M J 1968
Energy deposition in water by photons from point isotropic sources
MIRD Pamphlet No. 2, {\it J. Nucl. Med. Suppl.}  {\bf 1} 17-25

\item[] Cheung J Y C, Yu K N, Yu C P and Ho R T K 2001 
Dose distributions at extreme irradiation depths of gamma knife
radiosurgery: EGS4 Monte Carlo  calculations 
{\it Appl. Radiat. Isot.} {\bf 54} 461-5

\item[] Elekta 1992 {\it Leksell Gamma Unit-User's Manual} (Stockholm:
Elekta Instruments AB)

\item[] Elekta 1996 {\it Leksell GammaPlan Instructions for Use for
Version 4.0-Target Series} (Geneva: Elekta)

\item[] Hamad A, Mherat H 2005 (private communication)

\item[] Hubbell J H, Seltzer S M 2004
{\it Tables of X-Ray Mass Attenuation Coefficients and Mass
Energy-Absorption Coefficients (version 1.4)} {\tt
http://physics.nist.gov/xaamdi} (Gaithersburg: NIST)

\item[] Moskvin V, DesRosiers C, Papiez L, Timmerman R, Randall M and
DesRosiers P 2004 Monte Carlo simulation of the Leksell Gamma
Knife$^{\circledR}$: II. Effects of heterogeneous versus homogeneous
media for stereotactic radiosurgery 
{\it Phys. Med. Biol.} {\bf 49} 4879-95
 
\item[] Salvat F, Fern\'andez-Varea J M and Sempau J 2003
{\it PENELOPE - A code system for Monte Carlo simulation of
electron and photon transport} (Paris: NEA-OECD)

\item[] Solberg T D, DeMarco J J, Holly F E, Smathers J B and DeSalles
A A F 1998 Monte Carlo treatment planning for stereotactic
radiosurgery {\it Radiother. Oncol.} {\bf 49} 73-84

\item[] Wu A 1992 Physics and dosimetry of the gamma knife
{\it Neurosurg. Clin. N. Am.} {\bf 3} 35-50

\item[] Wu A, Lindner G, Maitz A, Kalend A, Lunsfond L D, Flickinger J
C and Bloomer W D 1990 Physics of gamma knife approach on convergent
beams in stereotactic radiosurgery {\it Int. J. Radiat. Oncol. Biol. 
Phys.} {\bf 18}, 941-9

\item[] Yu C and Shepard D 2003 Treatment planning for stereotactic
radiosurgery with photon beams {\it Technol. Cancer. Res. T.} {\bf 2}
93-104

\endrefs

\end{document}